# Prospects of odd and even harmonics generation by an atom in high-intensity laser field

A V Bogatskaya[1,3], E A Volkova[1] and A M Popov[1,2,3]

[1] D. V. Skobeltsyn Institute of Nuclear Physics, Moscow State University, 119991, Moscow, Russia
[2] Department of Physics, Moscow State University, 119991, Moscow, Russia
[3] P. N. Lebedev Physical Institute, RAS, Moscow, 119991, Russia

E-mail: alexander.m.popov@gmail.com

**Abstract.** New approach to study the spontaneous emission of the atomic system in the presence of the high-intensity laser field is used to study the process of harmonic generation. The analysis is based on the consideration of quantum system interaction with quantized field modes beeing in vacuum state, while the intense laser field is considered classically beyond the perturbation theory. The numerical analysis of the emission from the single one-electron one dimensional atom irradiated by the femtosecond laser pulse of Ti-Sa laser is discussed. It is demonstrated that not only odd but also even harmonics can be emitted if the laser field is strong enough. The origin of appearance of even harmonics is studied. The obtained results are compared with that found in the frames of semiclassical approach widely used to study the harmonic generation. It is found that semiclassical approach is inapplicable in the strong-field limit.

## 1. Introduction

Strongly nonequilibrium plasma produced in gases by high-intensity femtosecond laser pulses attracts the interest due to a number of specific electrodynamic properties that can be used for generation and amplification of radiation from X-ray and EUV to THz frequency band [1-13]. In particular, filamentation phenomenon can be very important in physics of the generation and (or) amplification of radiation [14].

Typically, the spontaneous emission from atoms and ions is not taken into account when strong-field dynamics of the atomic system is studied. Really, the value of vacuum modes of electric field strength is negligible in comparison with those in high-intensity femtosecond laser pulse. Typical time of allowed in the dipole approximation transitions for visible or UV radiation is in the range $10^{-6}$ -$10^{-8}$ s that many orders of value less than the duration of femtosecond pulse. Hence, it seems that there is no necessity to take such transitions into account. On the other hand, any nonlinear process in plasma starts from the spontaneous emission and at least at early stages of generation this emission should be taken into account. To overcome this difficulty the semiclassical approach [15] is widely used nowadays. This approach is based on the calculation of the response of the quantum system driven by the laser field using the average over the atomic wave function $\psi(\vec{r}, t)$ value of the dipole moment

$$\vec{d}(t) = -e \int \left| \psi(\vec{r}, t) \right|^2 \vec{r} \, d^3 r \qquad (1)$$

where $\psi(\vec{r}, t)$ describes the evolution of the atomic system in a laser field. If laser field is weak enough, i.e. its strength is much smaller than the atomic field strength value, the perturbation theory can be applied and one obtains the decomposition of the response over field strength powers [16]. If the field is strong and the perturbation theory to study the quantum dynamics is not applicable, the numerical solution of nonstationary Schroedinger equation for the quantum system in a laser field can be used. Such an approach to find the response and to study the contribution of different processes to filamentation was used in [17,18].

Similar approach was applied to study the atomic high - frequency response and high order harmonic generation (HHG) during last two decades [19-23]. In this case the intensity of emission was calculated as

$$I_\omega \sim \omega^4 \left| \vec{d}_\omega \right|^2 \qquad (2)$$

where $d_\omega$ is the Fourier transform of the dipole moment (1):

$$\vec{d}_\omega = \frac{1}{\sqrt{2\pi}} \int \vec{d}(t) \exp(-i\omega t) dt. \qquad (3)$$

Such an approach was found to be extremely fruitful, see, for example recent review [5] and references therein.

Nevertheless, the possibility to apply the semiclassical approach for study the harmonic generation and polarization response in the case of high-intensity laser field when the population of the initial (ground) state is depleted during the laser pulse action was questioned recently in [24]. It was demonstrated that the application of the semiclassical approach to study emission of the quantum system driven by high intensity laser field is generally in contradiction with quantum electrodynamical calculations. In [25] new approach to study the spontaneous emission of the atomic system in the presence of the high-intensity laser field was developed. This approach is based on the first order perturbation theory applied to the interaction of the atomic system dressed by the external laser field with a lot of quantized field modes in the assumption that initially all the modes are in a vacuum state.

Here we reformulate this approach for the velocity gauge and apply it to study the emission of the model single-electron atom driven by the femtosecond pulse of Ti-Sa laser ($\hbar\omega = 1.55$ eV). We find that odd harmonics of fundamental frequency are emitted by the atom only in rather weak external laser fields when the strong-field atomic dynamics can be studied in the frames of the quantum-mechanical perturbation theory. Beyond the applicability of the perturbation theory in the regime of effective atomic ionization both odd and even harmonics of the fundamental frequency are found to exist as a result of the electron bremsstrahlung in strong laser field.

The comparative analysis of the obtained data with those derived in the semiclassical approximation is performed. It is found that semiclassical approach that is widely used to study the high order harmonics generation fails in strong field limit when population of the ground state is depleted and essential ionization takes the place.

## 2. Theoretical model of interaction of the atomic system driven by the classical external field with quantized field modes.

We start the analysis of the spontaneous emission of the atomic system using the following Hamiltonian

$$H = H_0(\vec{r}, t) + H_f(\{a\}) + V(\vec{r}, \{a\}), \qquad (4)$$

where $H_0 = H_{at}(\vec{r}) + W(\vec{r}, t)$; $H_{at}(\vec{r})$ is the atomic Hamiltonian, and

$$W = -\frac{e}{mc} \vec{A}(t) \vec{p} + \frac{e^2 A^2(t)}{2mc^2} \qquad (5)$$

is the interaction of an atom with classical laser field in the velocity gauge in the dipole approximation, $\vec{A}(t)$ is the vector-potential of classical field, $\vec{p} = -i\hbar\nabla$ is the momentum operator, $H_f(\{a\})$ is the Hamiltonian of the set of field modes excluding laser field mode, $V(\vec{r},\{a\})$ stands for the interaction of an atomic electron with the quantized electromagnetic field, $\vec{r}$ is the electron radius-vector and $\{a\}$ is the set of quantized field mode coordinates.

Let us assume that we know the solution of the nonstationary problem for the atomic dynamics in the classical field

$$i\hbar\frac{\partial\psi(\vec{r},t)}{\partial t}=H_0(t)\psi(\vec{r},t) \qquad (6)$$

with the initial condition $\psi(\vec{r},t=0)=\phi(\vec{r})$ where $\phi(\vec{r})$ is a given stationary or unstationary state of the atomic discrete spectrum or continuum.

We will also suppose that at the initial instant of time all field modes are in the vacuum state $|\{0\}\rangle$. Then the solution of the general equation with the Hamiltonian (4)

$$i\hbar\frac{\partial\Psi(\vec{r},\{a\},t)}{\partial t}=\left(H_0(t)+H_f+V\right)\Psi(\vec{r},\{a\},t) \qquad (7)$$

and initial condition $\Psi(\vec{r},\{a\},t=0)=\phi(\vec{r})\times|\{0\}\rangle$ can be found by means of the perturbation theory. Wave function of zero-order approximation excluding interaction with the quantum field modes reads

$$\Psi^{(0)}(\vec{r},\{a\},t)=\psi(\vec{r},t)\times|\{0\}\rangle . \qquad (8)$$

We are going to find the solution of (7) in the form:

$$\Psi(\vec{r},\{a\},t)=\Psi^{(0)}(\vec{r},\{a\},t)+\delta\Psi(\vec{r},\{a\},t) \qquad (9)$$

with $\delta\Psi<<\Psi^{(0)}$.

For further analysis let us remind that initially we have vacuum in all field modes. Therefore in the first order of perturbation theory $\delta\Psi$ contains only one-photon excitations in a some field mode:

$$\delta\Psi(\vec{r},\{a\},t)=\sum_{k,\lambda}\delta\psi_{k\lambda}(\vec{r},t)\times\{0,0,....1_{k\lambda},0,...0,0\} \qquad (10)$$

Here $\delta\psi_{k\lambda}(\vec{r},t)$ is the electron wave function provided that one photon with wave vector $\vec{k}$ and polarization $\lambda$ has appeared.

As the interaction of the atom with quantized field can be written in a form

$$V(\vec{r},\{a\})=\sum_{k,\lambda}V_{k\lambda}=-\frac{e}{mc}\sum_{k\lambda}(\vec{e}_{k\lambda}\vec{p})a_{k\lambda} \qquad (11)$$

($a_{k\lambda}$ is the vector-potential operator of mode $\{k,\lambda\}$ and $\vec{e}_{k\lambda}$ is the polarization vector) for a given mode with one-photon excitation one has the problem

$$i\hbar\frac{\partial\delta\psi_{k\lambda}(\vec{r},t)}{\partial t}=H_0(t)\delta\psi_{k\lambda}(\vec{r},t)-\frac{e(\vec{e}_{k\lambda}\vec{p})}{mc}\times\frac{a_{norm}}{\sqrt{2}}\times\psi(\vec{r},t)\times\exp(i\omega_{k\lambda}t) \qquad (12)$$

($a_{norm}=\sqrt{4\pi\hbar c^2/(\omega_{k\lambda}L^3)}$, $L^3$ is normalization volume.) with the initial condition $\delta\psi_{k\lambda}(\vec{r},t=0)=0$.

Thus, we have the set of equations for atomic system evolution provided that one photon in the mode $\{\vec{k},\lambda\}$ has appeared. It is obvious, that the expression

$$W_{k\lambda}(t)=\int|\delta\psi_k(r,t)|^2 d^3r \qquad (13)$$

represents the probability to find a photon in the mode $\{\vec{k},\lambda\}$ as a function of time. Then the total probability to emit a photon of any frequency and polarization during the transition $f\rightarrow i$ is

$$W_{fi}(t)=\sum_{k,\lambda}W_{k\lambda}(t) . \qquad (14)$$

As the spectrum of field modes is dense, we can replace the sum in (14) by the integral over field modes. After the integration over angular distribution of photons and summation over possible polarizations the probability of spontaneous decay in the spectral interval $(\omega, \omega + d\omega)$ can be expressed in the form

$$W_\omega d\omega = \frac{L^3}{3\pi^2 c^3} \omega^2 d\omega \times W_{k=\omega/c,\lambda} , \qquad (15)$$

where $W_{k,\lambda}$ is given by (13). One should note, that the expression (15) do not depend on the normalization volume, as $W_{k,\lambda} \sim 1/L^3$ .

To provide more into the physics of spontaneous emission in the presence of strong laser field, wave functions $\delta\psi_{k\lambda}(\vec{r}, t)$ should be represented as a superposition of stationary states of the atomic Hamiltonian:

$$\delta\psi_{k\lambda}(\vec{r}, t) = \sum_n C_n^{(k\lambda)}(t)\varphi_n(\vec{r})\exp\left(-\frac{i}{\hbar}E_n t\right). \qquad (16)$$

The squared coefficients of the decomposition (16) $\left|C_n^{(k\lambda)}\right|^2$ have the sense to find the atom in the states $|n\rangle$ under the assumption that the emitted photon is in the definite mode $\{k, \lambda\}$. Further we will use these values to interpret the results of numerical simulation.

We would like to mention that the discussed problem can be formulated also in the length gauge [25]. In this case

$$W = -\vec{d}\vec{\varepsilon}(t) , \qquad (17)$$

where $\vec{d}$ is the dipole operator and $\vec{\varepsilon}(t) = -\frac{1}{c}\frac{d\vec{A}}{dt}$ is the electric field strength. However, it is known [15] that if the dipole approximation is valid, both gauges are equivalent to each other.

### 3. Numerical model

In this section we briefly describe the numerical model that was used to study spontaneous emission of the quantum system driven by high-intensity laser field. We study one-dimensional single-electron atomic system with Coulomb-screened potential [26]

$$V(x) = -\frac{e^2}{\sqrt{\alpha^2 + x^2}} \qquad (18)$$

with screening parameter $\alpha = 1.6165 a_0$, $a_0$ is Bohr radius. For such value of $\alpha$ the ionization potential is 12.13 eV which corresponds to ionization potential of xenon atom. The set of energy levels in such a xenon-like atom can be found in table 1.

**Table 1.** Xenon energy levels obtained in the numerical simulations for the potential (18)

| Principal quantum number, n | Energy (eV) |
|---|---|
| 1 | -12.134 |
| 2 | -5.910 |
| 3 | -3.457 |
| 4 | -2.220 |
| 5 | -1.550 |
| 6 | -1.135 |
| 7 | -0.870 |
| 8 | -0.685 |
| 9 | -0.555 |
| 10 | -0.457 |
| 11 | -0.385 |

Further we will discuss the set of calculations for an atom initially being prepared in the ground state n=1 and exposed to the radiation of the of Ti-Sa laser ($\hbar\omega$ =1.55 eV) with the trapezoidal sine-squared pulse with plateau of duration $t_p$ and smoothed sine-squared fronts of duration $t_f$, so that the total pulse duration was $\tau_p = t_p + 2t_f$. The parameters $t_p$ and $t_f$ were chosen to be equal of 2 and 10 optical cycles (OC).

According to the above-mentioned model we solve equation (6) for the evolution of atomic wave function self-consistently with the set of equations (12) for one-photon excitations in different quantized field modes. The procedure is briefly described in [25]. The frequency interval $\Delta\omega$ between quantized field modes was typically 0.02 of the fundamental frequency while the number of modes was 400-1200 in dependence of chosen value of laser intensity. The modeling was performed for the time interval twice exceeding the duration of pulse $\tau = 2\tau_p$. That allows one to distinguish spontaneous transitions that are possible without laser field from the stimulated transitions like stimulated Raman and Rayleigh type or stimulated bremsstrahlung when spontaneous photons appear only during the laser pulse action.

To compare the obtained spectra with those obtained in the semiclassical model semiclassical probability of emission in spectral interval $(\omega, \omega + d\omega)$ was also calculated

$$W_\omega^{(semi)} = \frac{4\omega^3}{3\hbar c^3} |d_\omega|^2 , \qquad (19)$$

where $d_\omega$ was taken from (1) and (3).

## 4. Discussion

We will start the discussion of the atomic dynamics in a strong laser field and harmonic generation from the data of ionization yield and probability to stay in the ground state in dependence on laser intensity (see table 2).

**Table 2.** Ionization yield and probability to stay in the ground state ($\hbar\omega$ =1.55 eV).

| Intensity (W/cm$^2$) | Ionization yield | Ground state population |
|---|---|---|
| 1E13 | 4,37274E-4 | 0,99927 |
| 2E13 | 0,00836 | 0,99086 |
| 3E13 | 0,08737 | 0,84413 |
| 4E13 | 0,28431 | 0,70324 |
| 5E13 | 0,52553 | 0,45724 |
| 6E13 | 0,45933 | 0,47493 |
| 8E13 | 0,58327 | 0,16376 |
| 1E14 | 0,94116 | 0,01716 |

First, we note that up to the intensity values $\leq 1 - 2 \times 10^{13}$ W/cm$^2$ the atom dominantly stays in its ground state. According to [24] that is the regime when semiclassical approach can be really used to study the spontaneous emission. Our simulations partially confirm this statement. The semiclassical calculations of atomic emission are presented at figure 1a. Due to the definite parity of atomic stationary states the peaks in spectra correspond to the odd harmonics only. Also three additional peaks are observed at energies ~6.22, 9.92 and 11.01 eV. These lines appear mainly in the after-pulse regime due to population of states $|n=2\rangle$, $|n=4\rangle$ and $|n=6\rangle$ and hence represent the oscillation of the dipole moment at consequent frequencies. These lines can be interpreted as spontaneous transitions to the ground state.

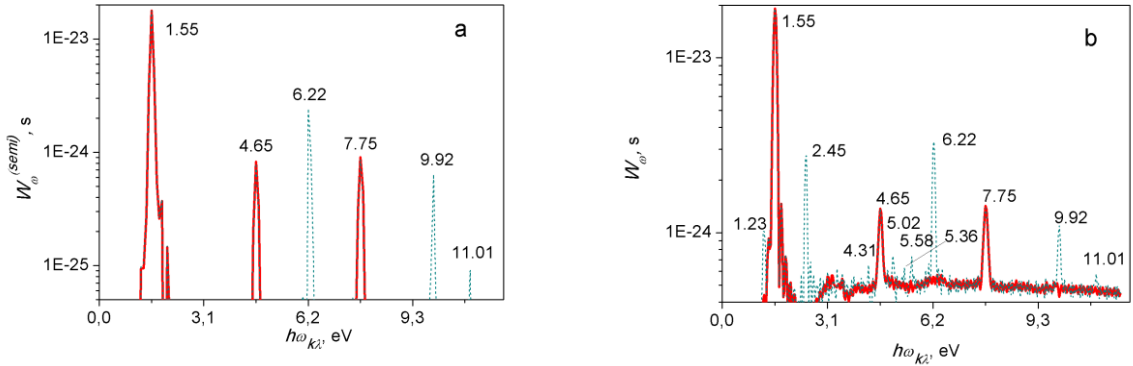

Figure 1. Semiclassical (a) and quantum-mechanical (b) spectra of spontaneous atomic emission for laser intensity $10^{13}$ W/cm$^2$. Solid curve is the emission spectrum by the end of the laser pulse, dash curve corresponds to the instant of time equal to two-pulse duration. Values near the peaks mean their position in eV.

Results of quantum-mechanical simulations of the atomic spontaneous emission for the same laser intensity are presented at figure 1b. One can see that there are two groups of lines in the spectra: one group corresponds to lines appearing during the laser pulse, while another one includes lines emitted dominantly in the after-pulse regime. First group of lines contains mainly the emission at fundamental frequency and its third and fifth harmonics. We could state that these harmonics are the result of Rayleigh $|n=1\rangle + \hbar\omega \rightarrow |n=1\rangle + \hbar\omega_{k\lambda}$ ($\omega = \omega_{k\lambda}$) and hyper - Rayleigh scattering when three of five laser photons are absorbed and one phonon is emitted. These lines are in rather good agreement with semiclassical calculations.

Besides lines corresponding to odd harmonics generation, several lines emerging in the after-pulse regime are also found to exist that partially do not appear in semiclassical calculations. Lines $\hbar\omega_{k\lambda} = 6.22, 9.92, 11.01$ eV correspond to $|n=2,4,6\rangle \rightarrow |1\rangle$ series of the spontaneous emission from levels excited during the laser pulse action. These lines are also observed in the semiclassical model. Nevertheless, one can distinguish a number of additional lines that do not exist in the semiclassical model. Among them there are lines with $\hbar\omega_{k\lambda} = 2.45, 4.31, 5.02, 5.36, 5.58$ eV that form the series $|n=3,5,7,9,11\rangle \rightarrow |2\rangle$. Line with $\hbar\omega_{k\lambda} = 1.23$ eV can be associated with the $|n=4\rangle \rightarrow |3\rangle$ transition. In both cases all these lines correspond to the transitions to unpopulated excited states and hence can not be observed in the semiclassical model [25].

The next point of our discussion is the spontaneous emission in the strong field limit when atom is dominantly ionized during the laser pulse. Semiclassical and quantum-mechanical spectra for the intensity value $10^{14}$ W/cm$^2$ are presented at figure 2. For this intensity value the probability of ionization is $\approx 0.94$ while the ground state is depleted up to $\approx 0.017$. In both cases one observes the harmonics of fundamental frequency with the plateau – like structure up to energies $\approx 3.17 U_p + I_i$ ($U_p$ is the ponderomotive potential and $I_i$ is the ionization potential of the atom). We would also like to mention the existence of spontaneous emission lines corresponding to transitions $|4\rangle \rightarrow |1\rangle$ and $|2\rangle \rightarrow |1\rangle$. Nevertheless, one can see that the results of quantum-mechanical calculations differ dramatically from those of semiclassical analysis. First, not only odd but also even harmonics are observed in the emission spectra in relatively low-energy part of the spectra $\hbar\omega_{k\lambda} \leq 10$ eV. In high-energy part of the spectra odd-order harmonics still dominate.

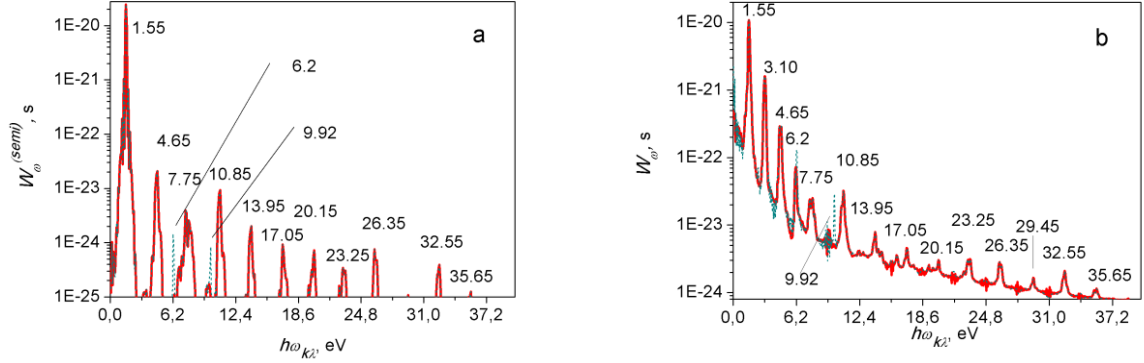

Figure 2. The same as figure 1, but for laser intensity $10^{14}$ W/cm$^2$.

To provide the insight into physical processes of harmonic emission in the regime of strong field ionization hereafter we analyze the distribution of population of ground state in dependence on harmonic frequency.

First, for strong laser field when the probability of ionization is high the peak with $\omega_{k\lambda} = \omega$ can have different physical nature. Besides the process of Rayleigh scattering it can arise from the near-free electron oscillations at the radiation frequency in the area of atomic potential. To distinguish these mechanisms the distribution of probabilities $\left|C_n^{(k\lambda)}\right|^2$ under the assumption that photon in the mode $\{k, \lambda\}$ with $\hbar\omega = 1.55$ eV was calculated. It was found that for $\hbar\omega_{k\lambda} = 1.55$ eV the probability to find the atom in ground state is $\approx 0.003$ and hence the bremsstrahlung dominates. Similar situation is also realized for a number of higher harmonics with $\omega_{k\lambda} = m\omega, \quad (m = 2-5)$. All of them come from the bremsstrahlung. As far as continuum states are degenerated and have different spatial parity for the same energy both odd and even harmonics can be emitted.

From our point of view oscillating and spreading of the ionized wave packet nearby the parent center is similar to the potential scattering in the presence of the strong external laser field when the quiver velocity is greater than the translational one. The spontaneous emission for such a bremsstrahlung regime was analyzed in [27]. Typical spectrum of the spontaneous emission obtained in [27] is presented at figure 3 and consists of a number of both odd and even harmonics. It is in a good qualitative agreement with our data of low-order harmonic emission.

As about high-order harmonics in the plateau regime still odd harmonics are dominantly emitted (see figure 2b). This part of harmonic spectra is in qualitative agreement with the semiclassical model. To explain this peculiarity the decomposition (16) of the wave function $\delta\psi_{k\lambda}(\bar{r}, t)$ was analyzed for field modes corresponding to $\omega_{k\lambda} = m\omega, \quad (m = 6 \div 24)$ in the intensity range $4 \times 10^{13} \div 10^{14}$ W/cm$^2$ when the plateau-like structure of harmonics is formed. It was found that both photorecombination to the ground state as well as bremsstrahlung contribute to odd harmonics, however namely the photorecombination provides the superior contribution to the plateau of odd harmonics. According to [27] we can conclude that the bremsstrahlung acts dominantly for the part of the spectra located before the plateau (see figure 2b) and as the result produces even harmonics there.

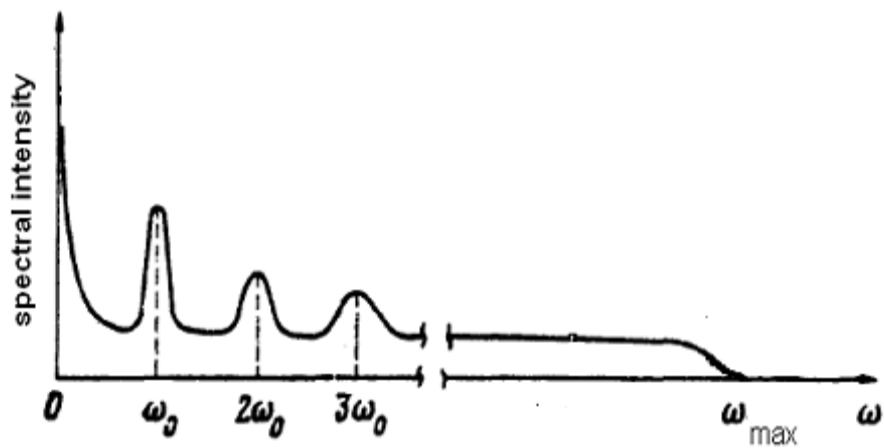

Figure 3. Spectral intensity of spontaneous bremsstrahlung in the field of an intense external wave.

It is also necessary to mention that the existence of the bremsstruhlung continuum was not observed in semiclassical calculations. It is found that quantum-mechanical calculation provides the intensity of harmonics and lines corresponding to transitions $|4\rangle \rightarrow |1\rangle$ and $|2\rangle \rightarrow |1\rangle$ up to order of value and larger then semiclassical ones. Both of these peculiarities result from the essential depletion of the ground state during the ionization process following by the suppression of transition to the ground state studied in the frames of semiclassical approach. As far as in the semiclassical approach the probability of photorecombination is proportional to the population of the final state [25], it means that this approach underestimates the efficiency of high-order harmonic generation in the strong-field limit.

It should be also noted that the emission of the single atom was analyzed in the present study. To produce the effective process of HHG one needs to analyze the emission from the atomic ensemble and the phase-matching effect. These problems are out of our consideration in this paper. Nevertheless, we can suppose that phase-matching first of all would suppress the emission arising from transitions between discrete atomic levels and in less degree those coming from bremsstrahlung and photorecombination.

## 5. Conclusions

Thus the analysis of the spontaneous emission of an atom in the presence of high-intensity laser field was performed. It is based on the considering of quantum system interaction with quantized field modes beeing in vacuum state while the intense laser field is accounted classically beyond the perturbation theory. It is demonstrated that not only odd but also even harmonics as well as lines associated with transitions between different discrete levels can be emitted if the laser field is strong enough. The origin of even harmonics is studied. It is explored that they result from the bremsstrahlung which becomes efficient in the regime of strong ionization. The obtained results are compared with those found in the frames of semiclassical approach. One can conclude that the semiclassical approach is inapplicable in the strong-field limit.

## Acknowledgments

This work was supported by the Russian Foundation for Basic Research (projects no. 15-02-00373, 16-02-00123). Numerical modeling was performed on the Lomonosov MSU supercomputer. Authors thank M.V. Fedorov for fruitful discussions.

## References

[1] Agostini P and Di Mauro L F 2004 *Rep. Prog. Phys.* **67** 813
[2] Winterfeldt C, Spielmann C and Gerber G 2008 *Rev. Mod. Phys.* **80** 117

[3] Krausz F and Ivanov M 2009 *Rev. Mod. Phys.* **81** 163

[4] Ganeev R A 2013 *Phys. Usp.* **56** 772

[5] Strelkov V V, Platonenko V T, Sterzhantov A F and Ryabikin M Yu 2016 *Phys. Usp.* **59** 425

[6] Kreβ M et al 2006 *Nature Phys.* **2** 327

[7] Gildenburg V B and Vvedenskii N V 2007 *Phys. Rev. Lett.* **98** 245002

[8] Wu H-C, Meyer-ter-Vehn J and Sheng Z-M 2008 *New J. Phys.* **10** 043001

[9] Silaev A A and Vvedenskii N V 2009 *Phys. Rev. Lett.* **102** 115005

[10] Bogatskaya A V and Popov A M 2013 *JETP Lett.* **97** 388

[11] Bogatskaya A V, Volkova E A and Popov A M 2014 *J. Phys. D.* **47** 185202

[12] Bogatskaya A V and Popov A M 2015 *Laser Phys. Lett.* **12** 045303

[13] Bogatskaya A V, Volkova E A and Popov A M 2016 *Laser Physics* **26** 015301

[14] Chin S L and Hu H 2016 *J. Phys. B* **49** 222003

[15] Fedorov M V 1997 *Atomic and Free Electrons in a Strong Light Field* (World Scientific, Singapore)

[16] Akhmanov S A and Nikitin S Yu 1997 *Physical Optics* (University Press, Oxford)

[17] Volkova E A, Popov A M and Tikhonova O V 2011 *JETP Lett* **94** 519

[18] Volkova E A, Popov A M and Tikhonova O V 2013 *J. Exp. Theor. Phys* **116** 372

[19] L'Huillier A, Lewenstein M, Salieres P, Balcou Ph, Ivanov M Yu, Larsson J and Wahlströmet C G 1993 *Phys. Rev. A* **48** R3433

[20] Kulander K C, Schafer K J and Krause J L , in *SILAP III*, (edited by B. Piraux) (Plenum Press, New York), 1993, pp. 95 – 110.

[21] Lewenstein M, Balcou Ph, Ivanov M Yu, L'Huillier A and Corkum P B 1994 *Phys. Rev. A* **49** 2117

[22] Becker W, Long S and McEver J K 1994 *Phys. Rev. A* **50** 1540

[23] Platonenko V T and Strelkov V V 1998 *Quantum Electron.* **25** 564

[24] Bogatskaya A V, Volkova E A, Kharin V Yu and Popov A M 2016 *Laser Phys. Lett.* **13** 0453014

[25] Bogatskaya A V, Volkova E A and Popov A M 2016 *EPL* **116** 14003

[26] Javanainen J, Eberly J and Su Q 1988 *Phys. Rev. A* **38** 3430

[27] Karapetyan R V and Fedorov M V 1978 *Sov. Phys. JETP* **48** 412